\documentclass{aa}
\thesaurus{03(03.09.6; 11.01.2; 11.03.3; 11.09.1 NGC 1275; 13.09.1; 13.09.4)}
\usepackage{graphics}
\begin{document}
%%%%%%%%%%%%%%%%%%%%%Title Page%%%%%%%%%%%%%%
\title {Near infrared imaging spectroscopy of NGC1275}

\author{Alfred Krabbe\inst{1,2} \and Bruce J.\,Sams\inst{2} III \and
Reinhard Genzel\inst{2} \and Niranjan
Thatte\inst{2} \and Francisco Prada\inst{3}}
\offprints{A. Krabbe; krabbe@dlr.de }
\institute{Deutsches Zentrum f\"{u}r Luft- und Raumfahrt, Institut f\"{u}r
Weltraumsensorik und Planetenerkundung, Rutherfordstr. 2,
12489 Berlin--Adlershof,
Germany \and Max-Planck-Institut f\"{u}r extraterrestrische Physik,
Giessenbachstra\ss e, 85748 Garching, Germany \and Centro Astronomico
Hispano--Aleman,
Almeria, Spain}
\date{Received / Accepted}
\maketitle

%%%%%%%%%%%%%%%Abstract Page%%%%%%%%%%%%%%%%%%
\begin{abstract}
We present H and K band imaging spectroscopy of the core regions of the
cD/AGN galaxy NGC\,1275.  The spectra, including lines from H{$_2$}, H,
$^{12}\mbox{CO}$ bandheads, \mbox{[Fe\,{\sc ii}]}, and \mbox{[Fe\,{\sc
iii}]}, are exploited to
constrain the star formation and excitation mechanisms in the galaxy's
nucleus.  The near-infrared properties can largely be accounted for by ionized
gas in the NLR, dense molecular gas, and
hot dust concentrated near the active nucleus of NGC\,1275.
The strong and compact H{$_2$} emission is
mostly from circumnuclear gas excited by the AGN and not from the cooling
flow.
The extended emission of late-type stars is diluted in the center by
the thermal emission of hot dust.
The line ratios \mbox{[Fe\,{\sc ii}]}/Br$\gamma$, as well as
\mbox{[Fe\,{\sc ii}]}/H{$_2$} can be explained with
X-ray excitation from the central AGN.


\end{abstract}

%%%%%%%%%%%%%%%%Begin Article%%%%%%%%%%%%%%%
\section{Introduction}

NGC\,1275 (Perseus A, 3C 84) at the center of the Perseus
cluster has been variously classified as an active galaxy, a merger, a
cooling flow galaxy, a BL Lac object, and an FR\,I radio jet source.
However, many of its physical characteristics --- from X-ray to radio
--- remain baffling.  NGC\,1275 exhibits the characteristics
of several source types: Seyfert (\cite{Seyfert:43}) included it in his
original list, Minkowski (\cite{Minkowski:57}) called it a galaxy
``collision''.  The dominant low velocity (LV) component
($\approx5200\mbox{\,km\,s$^{-1}$}$) is the giant cD elliptical
NGC\,1275, the other component (HV,
$\approx8300$\,km\,s$^{-1}$) is probably a nearby spiral galaxy
(at least partially) in front of the main galaxy (Haschick et al.
\cite{Haschick:82}).
Veron (\cite{Veron:78}) suggested that NGC\,1275 was a
\mbox{BL\,Lac}\, object, while other authors have suggested that it
was more compatible with LINERs (e.g. Norgaard-Nielsen et al.
\cite{Norgaard-Nielsen:90}).

NGC\,1275 has a powerful, nearly point like X-ray emitting nucleus
with a flux of
\mbox{\,L$_{\textup{X}}$}\,$\approx2\times$10$^{43}$\mbox{\,erg
s$^{-1}$}\, (Prieto \cite{Prieto:96}), which is embedded in an
extended halo of diffuse X-ray emitting intracluster gas.  The nucleus
of NGC\,1275 is displaced by about 25\arcsec\, from the center of
symmetry of the large scale ($\sim2$\arcmin) cluster gas halo, perhaps
due to its peculiar motion through the gas (B\"ohringer et al.
\cite{Bohringer:93}).  Temperature profiles of the X-ray halo suggest
that a ``cooling flow'' of\,\,
$\approx200$\mbox{\,M$_{\sun}$\,}\mbox{yr$^{-1}$}\, is condensing out
of the surrounding intracluster medium onto the galaxy (Fabian et al.
\cite{Fabian:81}, Cardiel et al. \cite{Cardiel:95}).  As in all cooling flow
galaxies, the ultimate fate of this purported inflowing material is
unclear, but visible band colors and Balmer absorption lines spectra
characteristic of A0 stars show that at most 10\% of the gas can form
stars with a standard Salpeter IMF (Romanishin \cite{Romanishin:86}).
NGC\,1275 is
unique among cooling flow galaxies in that it in fact shows evidence
of ongoing star formation.

Even by the standards of infrared luminous mergers, NGC\,1275
has very powerful, extended H{$_2$} emission (Fischer et al.
\cite{Fischer:87})
matched only by NGC\,6240 (van der Werf et al. \cite{vanderWerf:93}).  In
NGC\,1275,
this emission has been speculated to have its origin in the cooling
flow, although fast J-shocks, UV pumping, and X-ray heating via
supernovae and/or AGN also produce strong H{$_2$} emission.
Merging events are a likely mechanism for supplying NGC\,1275
with substantial (M\mbox{\lower0.6ex\hbox{$ \; \buildrel > \over
\sim \;$}}\, 3$ \times 10^{9}$\,M$_{\sun}$) quantities of
H{$_2$} in giant molecular clouds which are detected through mm
wave CO observations within the central regions (Inoue et al.
\cite{Inoue:96}).  CO
data also show two kinds of motion: a large (\mbox{\lower0.6ex\hbox{$
\; \buildrel > \over \sim \;$}} 5 \,kpc) scale rotation
which has been attributed to merging/cannibalism, and smaller scale
turbulence, attributed to the inflow of cooling flow gas
(Reuter et al. \cite{Reuter:93}).  Recent 2.6 mm CO interferometric
observations
reveal the majority of the gas to be in a ring structure of radius
$\approx4$\arcsec\, oriented in the East--West
direction.  The ring has a westerly extension up to
30\arcsec\, from the nucleus, and nearly $80\%$ of
the total CO is on the western side (Inoue et al. \cite{Inoue:96}).  Very deep
interferometric observations of the nucleus itself show no CO
absorption from very cold gas, which puts very tight limits on any
mass deposition from a cooling flow if the gas cools to nearly
3\mbox{\,K}\, (Braine et al. \cite{Braine-co1-0:95}).

\begin{table}
\begin{center}
\vspace*{0.3cm}
%{\sc Table 1} \\
%\vspace*{0.3cm}
%{\sc Observations of NGC\,1275 with the MPE 3D Imaging Spectrograph.} \\
\caption[Observations]{Observations of NGC\,1275 with the MPE 3D
imaging spectrograph.}
%\vspace*{0.3cm}
\begin{tabular}{llllll}
\hline
\hline
Date & Band & $t_{\textup{int}}$ & Seeing & Spectral/Flux &
m$_{\textup{V}}$ \\
  &  &  sec &   & Calibrator & \\
\hline
15/01/95  & K & 2400  & $1\farcs4$&  Mel 20497 (F6V)&  6.9\\
15/01/95  & H & 1100  & $1\farcs5$&   Mel 20497 (F6V)& 6.9\\
16/01/95  & K & 2000  & $ 1\farcs5 $&   PPM 45428  (F8V)& 4.1\\
16/01/95  & H & 1200 & $ 1\farcs6$&   Mel 20497 (F6V)& 6.9\\
21/01/95  & H & 2400  & $1\farcs1$ &    Mel 20497 (F6V)& 6.9\\
\hline
\end{tabular}
\end{center}
\label{tab-observations}
\end{table}

NGC\,1275 is a Fanaroff-Riley type I (\mbox{F--R\,\sc I}) radio source with
powerful
($\approx0.4$\,Jy at 22\,cm) radio lobes $\approx10$\,kpc long and
oriented at position angle (PA)160$^\circ$ which emanate from a
central core (Pedlar et al. \cite{Pedlar:90}).  The asymmetry of the radio
lobes and
their spectral indices suggest that we are looking nearly pole-on at
the central \mbox{F--R\,\sc I}\, radio engine (Pedlar et al.
\cite{Pedlar:90}, Levinson et al. \cite{Levinson:95}).  Within
the central 30\arcsec\, region, relativistic particles and the
magnetic field from the bipolar radio lobes produce a pressure, which
exceeds the thermal gas pressure and so must disturb any cooling flow
within that region (Pedlar et al. \cite{Pedlar:90}, B\"ohringer et al.
\cite{Bohringer:93}).  At even smaller
scales, VLBI imaging shows that the central core ($\approx1$\arcsec)
breaks up into many smaller components and a southern/northern jet
pair (Venturi et al. \cite{Venturi:93}, Vermeulen et al.
\cite{Vermeulen-counterjet:94}).  Recently, a new
component has been discovered only 55\,mas south of the central peak
(Taylor \& Vermeulen \cite{Taylor:96}).  Near infrared (NIR) annular
photometry shows that
the nuclear K-band light is dominated by non stellar emission which
could be either thermal emission from hot dust, or optically thin
synchrotron radiation (Longmore et al. \cite{Longmore:84}).  Recent {\em
HST}\, observations
in the UV show that any nuclear engine/star cluster must be $<$17\,pc
in size (Maoz et al. \cite{Maoz:95}).

Our goal in this work is to better constrain the many existing models
of NGC\,1275 with near infrared (NIR) imaging spectroscopy and spectral
synthesis modeling.  In dusty environments, NIR imaging has optical
depth penetration 10 times deeper than visible wavelengths, and NIR
spectroscopy is sensitive to a wealth of diagnostic lines which
describe the physical state of the gas and stars in the nucleus.
Throughout this paper we assume a scale of 344\,pc\,\arcsec$^{-1}$ (H$_0$ =
75\,km\,s$^{-1}$\,Mpc$^{-1}$, z = 0.0172).

\begin{figure*}
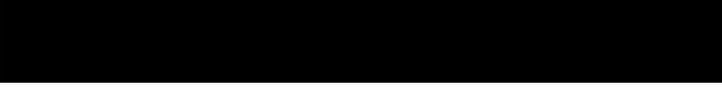

\resizebox{12cm}{8cm}{-}%\includegraphics{h1635_01.eps}}
\hfill
\parbox[b]{55mm}{
           \caption{K band spectrum of the central 3\arcsec\, of NGC\,1275.
	This aperture contains nearly all of the line flux.
	A linear baseline continuum of mean value
	$9.0\times 10^{-16}$ \, erg\,s$^{-1}$\,cm$^{-2}$\,\AA$^{-1}$\, has
been removed.
	}
           \label{fig-k-3secap-spectrum}}
\end{figure*}

 \begin{figure*}
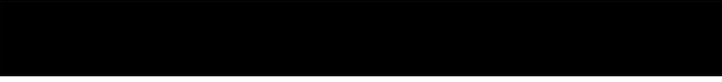

\resizebox{12cm}{7cm}{-}%\includegraphics{h1635_02.eps}}
\hfill
\parbox[b]{55mm}{
           \caption{H band spectrum of the central 3\arcsec\, of NGC\,1275.
	This aperture contains nearly all of the line flux.
	A linear baseline continuum of mean value
	$1.3\times 10^{-15}$\, erg s$^{-1}$\,cm$^{-2}$\,\AA$^{-1}$\, has
been removed.
	}
           \label{fig-h-3secap-spectrum}}
\end{figure*}

\section{Observations and data reduction}

We observed NGC1275 using the Max-Planck-Institut f\"{u}r
extraterrestrische Physik (MPE) imaging spectrometer 3D and the
Max-Planck-Institut f\"{u}r Astronomie (MPIA) tip-tilt image tracker
``CHARM'' on 1995 January 15, 16 and 21 at the 3.5m telescope of the
Max-Planck-Institut f\"{u}r Astronomie at Calar Alto, Spain.  The
CHARM instrument (McCaughrean et al. \cite{McCaughrean:94}) removes
atmospherically
induced image motion and produces stable images of increased spatial
resolution.  We locked CHARM on the bright point-like nucleus of
NGC1275, and corrected at 5\,Hz.  Under seeing conditions varying
from 1.0 to 1\farcs7, we achieve a final overall image quality of FWHM
1\farcs5 as determined by fitting to the point-like nucleus
(Maoz et al. \cite{Maoz:95}).  The ``3D'' instrument (Krabbe et al.
\cite{Krabbe:95}, Weitzel et al. \cite{Weitzel:96})
is an field imaging spectrograph which provides, during in a single
exposure, H or K band data cubes with spectral resolution of
R $=\lambda/\Delta\lambda = 1000~$and$~2000$.  The two spatial cube
dimensions are 16 pixels on a side, with pixel sizes selected to be
0.5\arcsec\, (giving an 8\arcsec$\times$8\arcsec\, field of view), while the
third dimension of 256 pixels is spectral.  The intrinsic spectral
sampling of the camera is with pixels of size $\lambda/$R, so by
dithering the spectral sampling by 1/2 pixel on alternate data sets,
we achieve fully Nyquist sampled spectra.  The spectral resolution,
achieved with the current reduction procedure, is 670, or 450\mbox{\,km
s$^{-1}$};
future software will produce full R = 1000 spectra after co-addition.
For extended objects, the efficiency gain over a long slit
spectrograph is substantial, as at any point within our
8\arcsec$\times$8\arcsec\, field, it is possible to obtain the full band
object spectrum.

The data (Table~\ref{tab-observations}) frames were interleaved
with identical sky exposures 60\arcsec\, away to the
West.  The source observations were preceded and/or followed
immediately by a set of spectral calibration observations of nearly
featureless stars of known brightness.  The few known
spectral features in the reference star spectra (e.g. Br$\gamma$\, in K band
and Br9 -- Br11 in H band) were later removed by interpolation.  A
temporally variable deep \mbox{CO$_{2}$} absorption near
2.01\,$\mu$m\, made accurate sky subtraction in this regime difficult.
In addition, the temporal variability of OH lines in the H band
produces spurious sky noise which dominates the background across the
band.

A run-down on typical 3D data reduction with the Groningen GIPSY
software is given in Weitzel et al. (\cite{Weitzel:96}).  Spectral
baselines were
calibrated using the color temperature of the calibrator stars and a
featureless Nernst glower.  The absolute wavelength calibration is
accurate to better than 1/4 pixel.  Individual data sets were properly
weighted, registered onto the central peak of NGC\,1275, and co-added.

 The K and H band spectra of the nucleus within a 3\arcsec\, aperture
 are presented in Figs.~\ref{fig-k-3secap-spectrum}
 and~\ref{fig-h-3secap-spectrum}, which present the enormous variety
 of emission and absorption lines.  The signal to noise in the K-band
 spectrum is $\approx$\,70 in the continuum.  For the H band, the
 signal to noise is $\approx$\,25 in the continuum.

\begin{table}
\begin{center}
%\vspace*{0.3cm}
%{\sc Table 2} \\
%\vspace*{0.3cm}
%{\sc Line Strengths in the Central $3\mbox{$^{\prime\prime}$}\,$ Aperture} \\
\vspace*{0.3cm}
\caption[Line strengths.]{Line strengths in the central
$3\mbox{$^{\prime\prime}$}\,$ aperture}
\begin{tabular}{lllll}
\hline
\hline
Species  & $\lambda^{\textup{a}}$ & Width $^{\textup{b}}$ & F$^{\textup{c}}$ &
$\sigma_{\textup{F}}$\\
\hline
FeII 1.534      &   1.5335   &   660        &   1.3 & 0.5\\
FeII 1.644      &   1.6439   &   526$^{\textup{d}}$    &   63 & 2.5\\
\mbox{H{$_2$}} 1--0 S(7)   &   1.7480   &   256          &   11 & 0.7\\
Br$\delta$   &    1.9446   &    263          &    3.1 & 0.5\\
\mbox{H{$_2$}} 1--0 S(3)   &   1.9576   &   301          &   42.5 & 2.5\\
\mbox{H{$_2$}} 1--0 S(2)   &   2.0338   &   225          &   14.4 &  0.7\\
HeI           &   2.0584   &   519          &   2.8  &  0.5\\
\mbox{H{$_2$}} 2--1 S(3)   &   2.0735   &   183          &   3.1  &  0.6\\
\mbox{H{$_2$}} 1--0 S(1)   &   2.1219   &   238          &   41.1 &  2.5\\
\mbox{H{$_2$}} 2--1 S(2)   &   2.1540   &   517          &   3.4  &  0.5\\
\mbox{Br$\gamma$}\,         &   2.1661   &   982         &   9.7 &  2.5\\
\mbox{H{$_2$}} 3--2 S(3)   &   2.008    &   207         &   1.3 &  0.5\\
\mbox{H{$_2$}} 1--0 S(0)   &   2.2231   &   328          &   10.0 &  0.7 \\
FeIII 2.244     &   2.243   &   504         &   2.2  &  0.5\\
\mbox{H{$_2$}} 2--1 S(1)    &   2.2473   &   754          &   7.7  &  0.6\\
\hline
\end{tabular}
\end{center}
$^{\textup{a}}$  \mbox{\,$\mu$m}\, observed corrected for $z=0.0172$ \\
$^{\textup{b}}$ \mbox{\,km s$^{-1}$}\, corrected for instrument resolution
of 450 \mbox{\,km s$^{-1}$}.  Width
error typically 30 -- 50 \mbox{\,km s$^{-1}$}\,\\
$^{\textup{c}}$ $10^{-15}$ erg \mbox{\,s$^{-1}$}\, cm$^{-2}$\\
$^{\textup{d}}$ Best fit of single Gaussian.  Two component fit of $1345
\pm 120
\mbox{\,km s$^{-1}$}\,$ and
$280 \pm 22 \mbox{\,km s$^{-1}$}\,$ is superior. See text.
\label{tab-line-strengths}
\end{table}

\section{Nature of the continuum emission}
\label{sec-cont-emission}
\begin{figure}
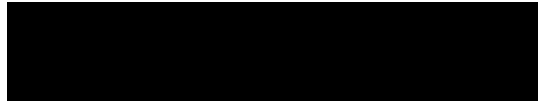

\resizebox{\hsize}{9cm}{-}%\includegraphics{h1635_03.eps}}
\caption{The K band continuum of NGC\,1275.
	Orientation: N up, E left. FWHM $1\farcs6$.
	Field of view 8\arcsec$\times$8\arcsec.
	Contours are at 1.25, 2.5, 5, 10, 20, 30\ldots\% of
	peak flux of $8.5 \times 10^{-15}$\, erg
s$^{-1}$\,cm$^{-2}$\,arcsec$^{-2}$.
	}
\label{fig-k-cont-map}
\end{figure}

Fig.~\ref{fig-k-cont-map} represents the true, line emission free, K
band continuum structure of NGC\,1275.  Its peaking at the nucleus
confirms that the bulk of the emission arises in the central 300
\,pc.  The slight south-east north-west extension, almost along the
radio PA of 160$^\circ$ (Pedlar et al. \cite{Pedlar:90}), is 2$\sigma$
significant.

We analyzed the the continuum spectrum with respect to contribution
from stars, the active galactic nucleus, hot dust from the
circumnuclear environment, and free-free emission.  K and M stars
(T$_{\textup{eff}}\,\approx4000$), dominating the K band stellar emission, show
strong CO absorption features in their more or less BB spectrum which
can be used to constrain the fraction of total light originating in
stars.  NIR spectra from AGN show a power law spectrum of the form
I$_{\lambda} \propto \lambda^{\alpha}$ where $\alpha$ varies between
$-0.5$ and $-1.5$ (Laor \& Draine \cite{Laor:93}, Acosta-Pulido et
al. \cite{Acosta-Pulido:90}).  Hot dust in this
context is mainly referred to as being reprocessed radiation from AGN
and, to a minor extent, from starburst activity.  It's spectrum is
that of a black body multiplied by a power-law emissivity of the form
$\epsilon_{\textup{dust}} \propto \lambda^{-1}$.  From these contributions a
synthetic galaxy spectrum including extinction effects has been
modelled and (least square) fitted to the observed continuum spectrum.
Details of the procedure are given in the appendix, contrains on the
parameters are being discussed below, and the results are presented in
Sect.\,\ref{sec-decomposition}.

\subsection{Physical constrains on the continuum emission}
\label{sec-constrains}
The size of the search space was reduced by placing constrains on the total
extinction, Galactic extinction, free-free contribution, dust temperature,
and stellar populations.

\emph{i}) The Galactic extinction toward NGC\,1275 is A$_{\textup{B}}$ =
0.7, or
A$_{\textup{V}}$ = 0.53, and there is an additional component of extinction
intrinsic to NGC\,1275 itself.  From UV measurements
we have E(B-V) = 0.2 -- 0.3 (Maoz et al. \cite{Maoz:95}, Levinson et al.
\cite{Levinson:95}), which
both suggest that A$_{\textup{V}} \approx 1.0$ at the nucleus.  Integrated
over
the central stellar area, H--K is 0.15 redder than typical ellipticals
(Longmore et al.  \cite{Longmore:84}), which requires A$_{\textup{V}}
\approx 2.4$.  Hence we
constrain the V band extinction to A$_{\textup{V}} < 2.0$, or
A$_{\textup{K}} < 0.2$
(Rieke \& Lebofsky \cite{Rieke:85}).  Since we know that the central light
is likely to
be dominated by the AGN (Longmore et al. \cite{Longmore:84}), we choose to
use a
screen model of extinction there, while farther out at distances
\mbox{\lower0.6ex\hbox{$ \; \buildrel > \over \sim
\;$}}\,500\,pc we will use a mixed stars/gas model.

\emph{ii}) The free-free continuum can be obtained from the Br$\gamma$ line
flux using case B recombination and assuming T = 10000\,K.
(Joy \& Lester \cite{Joy:88}) have shown that this method gives consistent
results
compared to a NIR multicolor continuum decomposition (Joy \& Harvey
\cite{Joy:87})
and radio measurements (Wynn-Williams \& Becklin \cite{Wynn:86}).  Using
the Br$\gamma$ line flux
from Table ~\ref{tab-line-strengths} we can thus constrain the
free-free contribution to be about $2\%$ of the underlying continuum.
There is some suggestion that the free-free emission could be higher
in NGC\,1275 because the central plasma is constantly being replenished
with free electrons via the cooling flow gas.  However, if there were
a substantial population of such ions, they would contribute to the
scattering of the central source at the 1\% level (Sarazin \& Wise
\cite{Sarazin:93}),
which would in turn produce a blue excess.  If the 1\% scattering is
responsible for the observed nuclear blue excess of $\sim 10^{44}$
\,erg\,s$^{-1}$\, (Crawford \& Fabian \cite{Crawford:93}), then the central
source would require
an X-ray/UV luminosity of roughly 10$^{46}$\,erg\,s$^{-1}$, which is
\,\,$\approx 600$ times the point X-ray luminosity.  Hence we set the
free-free emission to a typical value of $<3\%$ of the total
continuum, or sometimes ignore it completely, as our fits constrain
the relative contributions to only $\approx \pm 10\%$.  Had we not
done so, the free-free and AGN emission would have traded off
influence with each other, as they are both power laws of similar
slopes.

\emph{iii}) Hot dust has turned out to be very common in the nuclear
regions of AGNs (Granato et al. \cite{Granato:97}, Maiolino et al.
\cite{Maiolino:95}) and, to some extent,
in starburst galaxies as well (see e.g. B\"oker et al. \cite{Boeker:98}).  It
surrounds the nucleus forming a torus which absorbs radiation from the
AGN and reemits it in the mid-infrared.  Since hot dust too close to
the AGN will start to sublime, the maximum dust temperature is set by
the sublimation temperature of the dust material.  Silicate and
graphite grains, which are the main constituents of interstellar dust
(Laor \& Draine \cite{Laor:93}), will sublime at temperatures of about
1400K and
1750K resp., setting the limit for the the dust temperature at about
1500K at the inner edge of the dust torus.  The minimum radius for
dust grains to survive is \mbox{\lower0.6ex\hbox{$ \; \buildrel > \over \sim
\;$}}\,0.2L$_{46}^{1/2}$ \,pc
(Laor \& Draine \cite{Laor:93}), where L$_{46}$ is the central source
bolometric
luminosity in units of $10^{46}$\, erg\,s$^{-1}$.  For NGC\,1275 with
L$_{46}$ = 0.04 (Levinson et al. \cite{Levinson:95}), this is
\mbox{\lower0.6ex\hbox{$ \; \buildrel > \over \sim
\;$}}\, 0.05 \,pc.  The
dust temperature observed will be a mixture of dust at different
temperatures along the line of sight weighted with their column
density distribution.  Thus the effective temperature will be lower
and, e.g. for NGC 1068, in the range of 700\,K (Thatte et al.
\cite{Thatte:97}).  As
NGC\,1275 is a \mbox{F--R\,\sc I}\, type radio source, we are probably
looking nearly
pole on at the central source, and hence see the largest possible
fraction of the hot inner surfaces of the clouds which make up an
obscuring torus (Pier \& Krolik \cite{Pier:92}), suggesting that we limit the
temperature of the dust in our simulations to a relatively high value
of 1000\,K.

\emph{iv}) The NIR flux of essentially all integrated galaxy
populations is dominated by giants and, to a lesser extent, by
supergiants (Oliva et al. \cite{Oliva:95}).  We rule out \mbox{K5\,\sc V}\,
stars because for
all reasonable stellar populations they are too faint to contribute
substantially to the overall H and K band fluxes.  As an underlying
stellar population we choose \mbox{K5\,\sc III}\, stars because they
dominate, to
first order, the light of nearly all galaxies at H and K bands
(Frogel et al. \cite {Frogel:78}, Oliva et al. \cite{Oliva:95}).  This is
due to the fact that their
effective temperatures of \,\,$\approx 4000$\,K have a black body
emission peak in those bands, and because they are typically 200 times
brighter than \mbox{K5\,\sc V}\, stars with similar effective temperatures.
Digitized libraries of NIR stellar spectra are available in the K band
\cite{Kleinmann:86}, so we used the spectrum of the \mbox{K5\,\sc III}\, star
$\gamma$\,Dra as our template I$_{\lambda}^{\textup{STAR}}$ in the K band.

In the H band, no such digitized stellar library was publicly
available at the time of this data reduction, so we constructed a
continuum spectrum appropriate to the stellar population of NGC\,1275
by summing the contributions of many black body curves of different
temperature scaled by the fractional number of stars having that
temperature.  This produces a continuum spectrum without any lines.
The H band CO absorption features are relatively weak, so they would
have contributed little to the fitting anyway.  The galaxy population,
necessary for this procedure, was modeled with the stellar population
evolution code STARS (Sternberg \& Kovo, in preparation).  Given a set of input
parameters it determines the binned number density function of stars
in the HR-diagram.  Summing the series of appropriately scaled black
body curves at the various temperatures results in a H-band continuum
that approximates the underlying H band continuum of NGC\,1275, if the
input parameters to STARS are varied adequately.  The composite
spectrum is then normalized to the same wavelength as the K band
stellar library spectrum in order to produce the final H and K band
underlying stellar spectrum, I$_{\lambda}^{\textup{STAR}}$.  All K spectra
are normalized to a
wavelength of 2.27\,$\mu$m, and all H spectra are normalized to 1.57\,$\mu$m,
which we chose because the spectra are free of lines there.

\begin{figure}
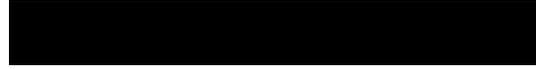

\resizebox{\hsize}{6cm}{-}%\includegraphics{h1635_04.eps}}
\caption{Spectral fitting to the NGC\,1275 spectrum in an aperture
	of 1\arcsec\, centered on the nucleus.  Best fit is achieved with a
	relative stellar contribution of 0.0, an AGN
	power-law (see text) contribution of 0.5, a hot dust
	contribution of 0.5 at 700 K, and a K-band extinction of 0.05
	(screen model).  The plot shows (in decreasing order from the top)
	a) the raw spectral data; b) the final derived model fit to the
	continuum; c) the statistical weights given to each data point;
	d) the derived power-law component of the
	model; e) the derived dust component of the model; f) the residuals
	between model and data.  Note that because we are fitting the
	continuum, not the lines, the weights in regions of strong line
	emission have been set to zero (cf Eq.~(\ref{eqn-chisqr})).
	The spectra are all normalized at
	$\lambda$ = 2.27\,$\mu$m, so the relative contributions of the various
	components can be read off at that wavelength.}
\label{fig-specfit-hk-0}
\end{figure}

\subsection{Decomposing the continuum emission}
\label{sec-decomposition}
In the central 1\arcsec\, (0.3\,kpc) region we expect that the contribution
of the AGN
overwhelms the the stellar population.  This is indeed the case, as the
continuum emission from the center of NGC\,1275 is well modeled by the
light of \mbox{K5\,\sc III}\, or \mbox{K5\,{\sc I}b}\, stars heavily
diluted by emission from
hot dust and AGN typical power-law emission (Fig.~\ref{fig-specfit-hk-0}).
The best fitting values are
$c_{\textup{STAR}} = 0.0 \pm 0.1$,
$c_{\textup{PL}} = 0.5\pm 0.1$,
$c_{\textup{HD}} = 0.5\pm 0.1$,
$T_{\textup{d}} = 700\pm 200$,
$\tau_{\textup{K}} = 0.05 \pm 0.1$,
where the $c_{\textup{i}}$ represent fractional contributions and were the
errors are those derived from the estimated systematic
errors in the spectra and from multiple trials with slightly varied
weights and initial guess parameters.  The relative insignificance of
spectral features in the underlying stellar population shows that the
fitting process at the nucleus is dominated by the power laws, and
that we are unable to distinguish between the various possibilities of
underlying populations.  The relative contribution of power-law emission is
not particularly sensitive to the adopted AGN spectral index (-0.5
through -1.5, see appendix), because
its main function is to dilute the very steep slope of the stellar
continuum (very nearly a black body).

\begin{figure}
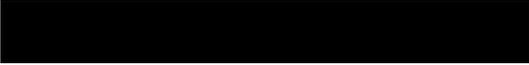

\resizebox{\hsize}{5.8cm}{-}%\includegraphics{h1635_05.eps}}
\caption{Spectral fitting to the NGC\,1275 spectrum in a 1\arcsec\, wide
annulus
	of radius 3\arcsec\, centered on the nucleus.
	Best fit is achieved with a relative stellar contribution of 0.6, an
	AGN power-law (see text) contribution of 0.4, a hot dust
	contribution of 0.0, and a K-band extinction of 0.01
	(mixed stars and gas model).}
\label{fig-specfit-k-3}
\end{figure}

In contrast, the same fitting process applied to a  1\arcsec\, wide
annulus 3\arcsec\, (0.9\,kpc) from the nucleus yields completely different
result.  Because
the CO absorption bandheads are clearly observed in the K band
spectrum (they are much less diluted), we fit
this region only using the K band data.  The fit is excellent
(Fig.~\ref{fig-specfit-k-3}), and yields the following contributions:
$c_{\textup{STAR}} = 0.6\pm 0.1$,
$c_{\textup{PL}} = 0.4 \pm 0.1$,
$c_{\textup{HD}} = 0.0\pm 0.1$,
$\tau_{\textup{K}} = 0.01 \pm 0.1$, where we have chosen to apply a mixed
stars/gas model of extinction rather than a simple screen.
Again, we assume a \mbox{K5\,\sc III}\, star dominated underlying stellar
population.  Note that the AGN contribution could also be interpreted
as free-free emission at this distance from the nucleus, where we do
not expect any direct or scattered contribution from the AGN.  By the
time we reach 4\arcsec\, (1.2\,kpc) from the nucleus, the stellar
component dominates completely, and we find
$c_{\textup{STAR}} = 0.90\pm 0.1$,
$c_{\textup{PL}} = 0.1 \pm 0.1$,
$c_{\textup{HD}} = 0.0\pm 0.1$,
$\tau_{\textup{K}} = 0.01 \pm 0.1$.  Hence the behavior of the the various
components is as expected from a central AGN dominated source.

\section{The H{$_2$} emission}

The H{$_2$} emission in NGC\,1275 is exceptionally luminous
(Fischer et al. \cite{Fischer:87}) even if compared with other powerful
H{$_2$} emitting
galaxies.  The excitation of H{$_2$} in AGNs is usually explained to be a
mixture of shocks (Brand et al. \cite{Brand:89}), UV photons
(Sternberg \& Dalgarno \cite{Sternberg:89}),
and X-rays (Draine \& Woods \cite{Draine:91}).  The shocks can be caused by
galaxy
merging or by supernovae, the UV radiation could come from OB star
associations (Puxley et al. \cite{Puxley:90}), while only an AGN with a
black hole
can produce the quantities of X-ray radiation needed to excite the H{$_2$}
seen in Seyfert nuclei.

\subsection{Spatial distribution of H{$_2$} emission}
\label{sec-h2-distribution}
\begin{figure}
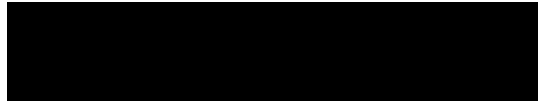

\resizebox{\hsize}{9cm}{-}%\includegraphics{h1635_06.eps}}
\caption{Line map of the H{$_2$} 1--0 S(1) transition.
	The emission is almost completely symmetric, with FWHM of 1\farcs6.
	The other H{$_2$} transitions show similar structure and extent.
	Contours are at 2.5, 5, 10, 20, 30\ldots\% of peak flux of
	$2.1\times 10^{-14} \mbox{\, erg s$^{-1}$ cm$^{-2}$ arcsec$^{-2}$}$.}
\label{fig-h2-line-map}
\end{figure}

\begin{figure}
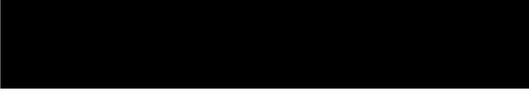

\resizebox{\hsize}{8.5cm}{-}%\includegraphics{h1635_07.eps}}
\caption{Azimuthally averaged profile of the H{$_2$} 1--0 S(1) emission.
	There is a very low level extended component to the emission, which
	reaches a few percent of the peak flux at r = 4\arcsec.}
\label{fig-h2-line-azim-avg}
\end{figure}

The H{$_2$} line emission (Fig.~\ref{fig-h2-line-map}) is strongly
dominated by an unresolved source centered on the nucleus surrounded
by weak extended emission extending a few arcseconds west.
The fact that the primary exciting source is not distributed over
scales \mbox{\lower0.6ex\hbox{$ \; \buildrel > \over \sim
\;$}}\,300\,pc suggests that most of the H{$_2$} emission is
associated with the circumnuclear environment and probably the active
galactic nucleus and not, as speculated before (Inoue et al.
\cite{Inoue:96}) with
the cooling flow.  The low level extended emission reach $\approx 2\%$
of the peak flux at our limiting distance of 5\arcsec\, or 3.5 \,kpc
(Fig.~\ref{fig-h2-line-azim-avg}).  The total H and K band H{$_2$}
emission amounts to $1.3 \times 10^{-13}$ \,erg\,s$^{-1}$\, cm$^{-2}$ (see
Table~\ref{tab-line-strengths}) or, for a distance of 70\,Mpc, a
total luminosity of L$_{\textup{H}{_2}} = 8 \times 10^{40}$\, erg\,s$^{-1}$.

\subsection{Excitation of the H{$_2$} emission}
\label{sec-h2-excitation}
The high signal-to-noise in our spectra allow us to check the effective
excitation temperature from the column density of
molecules in each quantum state (see e.g. Beckwith et al. \cite{Beckwith:78}):
\begin{center}
\begin{equation}
	~~~~~~~~~N_{col} = 4\pi I/A_{ul}h\nu
	\label{eqn-line-excitation-temp}
\end{equation}
\end{center}
where $I$ is the surface brightness in the line and $A_{\textup{ul}}$ is the
Einstein A coefficient.  The surface brightnesses have been dereddened using
A$_{\textup{V}} = 2.0$ (Sect.\,\ref{sec-constrains}) as a realistic value.
Screen and mixed
extinction models (McLeod et al. \cite{McLeod_al:93}) for the dereddening
gave  flux
correction between 10\% and 20\%. Since the appropriate model
was uncertain, the correction was averaged between screen and mixed model
and the uncertainties propagated in the errors.

In thermal equilibrium, the line ratios should be given solely
by the number of atoms in each quantum state and their spontaneous
emission coefficients if the H{$_2$} is optically thin.  From
Table~\ref{tab-h2-col-dens} we conclude that this is indeed the case.
The column densities are always lower than
N$_{\textup{col}}$(H{$_2$}\,1--0 S(1))$ = 9.6
\times 10^{15}$\,cm$^{-2}$.  Summing over all the observed H{$_2$} lines
gives N$_{\textup{col}}$(H{$_2$}$^{\textup{hot}}) \approx 3 \times 10^{16}$
\,cm$^{-2}$.  If we
assume that this gas is uniformly distributed within the emitting
region, then the total mass of contributing H{$_2$} would be $\sim 1000$
\,M$_{\sun}$.

\begin{figure*}
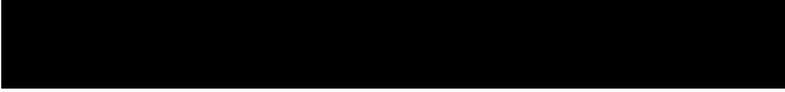

\resizebox{13cm}{8.5cm}{-}%\includegraphics{h1635_08.eps}}
\hfill
\parbox[b]{45mm}{
           \caption{Upper level column density per mode of observed H{$_2$}
lines
	against upper level energy in $\mbox{\,K}$. A
	slope in this diagram is equivalent to an excitation temperature.
	}
           \label{fig-h2-exite-temp}}
\end{figure*}

In Fig.~\ref{fig-h2-exite-temp} we plot the natural log of
N$_{\textup{col}}$
per mode against the energy of the upper level given as a temperature
(compare with Kawara \& Taniguchi \cite{Kawara:93}).  The error bars
include the errors
given in Table~\ref{tab-observations} and the dereddening.  The
inverse slope is the effective excitation temperature T$_{\textup{ex}}$ of the
gas and corresponds to the kinetic temperature if local thermal
equilibrium (LTE) prevails.

Depending on the conditions within a molecular cloud, the initial
ortho/para ratio of 3 of molecular hydrogen at formation may evolve
and eventually drop to lower values (Martini et al. \cite{Martini:97}).
Exchange
reactions with H and H$^+$ (Flower \& Watts \cite{Flower:84}) or a low
temperature
environment (T \mbox{\lower0.6ex\hbox{$ \; \buildrel < \over \sim
\;$}}\, 200\,K; Burton et al. \cite{Burton:92}) can even lower the
ratio to below 2.0 (see e.g. Hora \& Latter \cite{Hora:96}).  A high
temperature
environment in turn will keep the ortho/para ratio at 3.  In a
photodissociation region (PDR), however, freshly exposed H{$_2$} may keep
the ortho/para ration down (Chrysostomou et al.
\cite{Chrysostomou:93}) although the
environment is hot.  In Fig.~\ref{fig-h2-exite-temp} the column
densities have been divided by the statistical weights corresponding
to their states which include the spin degeneracies g$_{\textup{s}} = 1$ for
even (para) and g$_{\textup{s}} = 3$ for odd (ortho) J. If the ortho/para
ratio
in the molecular hydrogen is 3:1 then the ortho and para column
densities should lie along a line.  For the 1--0 transitions, were we
have enough data points, this is indeed the case.  This demonstrates
that the either the molecular hydrogen has just been formed or that
the H{$_2$} has been kept in a high temperature environment.

The distribution of data in Fig.~\ref{fig-h2-exite-temp} shows that
the H{$_2$} gas is nearly isothermal.  Below 8000\,K upper level energy,
the excitation temperature is T$_{\textup{ex}} \approx 1500$\,K, but above
8000\,K, T$_{\textup{ex}}$ does only vary between $\approx 2600$\, K and
$\approx
2900$\, K.  If no extinction correction is applied, the temperatures come out
to be 1450\,K, 2400\,K, and 2650\,K resp., which is still in agreement with
the 1$\sigma$ errors given in Fig.~\ref{fig-h2-exite-temp}.  All date
(except for 2--1\,S(3), see below) out to level energies of
$\approx20000$\,K lie along an only slightly curved line.  In
particular, the column density per mode of 1--0\,S(7) is close to those
of 2--1\,S(1) and 2--1\,S(2), indicating that the rotational and
vibrational temperatures are about the same and the fluorescent
excitation of the gas is rather small (Draine \& Bertoldi
\cite{Draine-Bertoldi:96}).
The proposed likely mechanism to populate the levels of this hot H{$_2$}
gas with T$_{\textup{ex}} \le 2700$\, K is collisional excitation in a post
shock
environment, which has been observed in many galactic sources
(see e.g. Fischer et al. \cite{Fischer:87}, Gredel et al.
\cite{Gredel:94}, Wright et al. \cite{Wright:96}).  However, at column
densities prevailing in the central 300\,pc of NGC\,1275, a PDR scenario
cannot be completely excluded.  In dense PDRs, the vibrational ladder
is thermalized while the rotational ladder in a given vibrational
state may be out of thermal equilibrium (see
e.g. Timmermann et al. \cite{Timmermann:96}).  The fact that we do not see
gas of
temperature above 5000\,K (Wright et al. \cite{Wright:96}) may, however,
indicate
that the contribution of nonthermal processes (UV pumping) to the H{$_2$}
excitation is rather moderate.

\begin{table}
\begin{center}
%\vspace*{0.3cm}
%{\sc Table 4} \\
%\vspace*{0.3cm}
%{\sc Column Density and Temperature Analysis of \mbox{H{$_2$}}} \\
\vspace*{0.3cm}
\caption[Temperature analysis]{Column density and temperature analysis of
\mbox{H{$_2$}}}
\begin{tabular}{lllll}
\hline
\hline
Line  & A$_{\textup{ul}}^{\textup{a}}$ & N$_{\textup{col}}^{\textup{b}}$ &
$ln(N_{\textup{col}}/g_{\textup{u}})$ &T$_{\textup{ex}}^{\textup{A}}$\\
\hline
1-0 S(0)   &   2.5(-7)   & 3.8(15) & 43.47    &   6472 \\
1-0 S(1)   &   3.5(-7)   & 1.1(16) & 43.09    &   6951 \\
1-0 S(2)   &   4.0(-7)   & 3.2(15) & 42.72    &   7585 \\
1-0 S(3)   &   4.2(-7)   & 8.7(15) & 42.42    &   8365 \\
1-0 S(7)   &   3.0(-7)   & 2.9(15) & 40.79    &  12817 \\
2-1 S(1)   &   5.0(-7)   & 1.5(15) & 41.10    &  12551 \\
2-1 S(2)   &   5.6(-7)   & 5.6(14) & 40.98    &  13150 \\
2-1 S(3)   &   5.8(-7)   & 4.9(14) & 39.53    &  13887 \\
3-2 S(3)   &   5.6(-7)   & 2.0(14) & 38.66    &  19089 \\
\hline
\end{tabular}
\end{center}
$^{\textup{a}}$ Turner et al. \cite{Turner:77} \\
$^{\textup{b}}$ cm$^{-2}$; H{$_2$} flux dereddened with A$_{\textup{V}}=2.0$ \\
\label{tab-h2-col-dens}
\end{table}

The excitation conditions in NGC\,1275 can be investigated in more
detail by examining the ratio of two ortho or para lines and put them
in context.  The remaining uncertainty introduced by the intrinsic
ortho-para ratio of the gas clouds can thus be avoided (Mouri
\cite{Mouri:94}).
Locating these ratios in a ortho-para line ratio
plane provides an independent guide for distinguishing between the
relative contribution of X-ray heating, shocks, and UV pumping
excitation of H{$_2$} gas in starburst and AGN galaxies
(Fig.~\ref{fig-mouri-12}).  As non thermal processes (e.g. UV pumping
in low density gas) excite high vibrational states more easily than
the thermal process, the ratio of two ortho transitions such as
R$_{\textup{o}}$
= H{$_2$}\,2--1\,S(1)/H{$_2$}\,1--0\,S(1) gives an ortho-para balance
independent
measure of the relative contribution of thermal and non-thermal
processes.  The same holds for R$_{\textup{p}}$ =
H{$_2$}\,1--0\,S(2)/H{$_2$}\,1--0\,S(0).  The
expected value of R$_{\textup{o}}$ in the case of pure non thermal excitation
under nearly all realistic conditions is R$_{\textup{o}} = 0.56$
(Black \& van Dishoeck \cite{Black:87}), while thermal UV excitation gives
a value of
R$_{\textup{o}}$ near 0 (Sternberg et al. \cite{Sternberg:87}).  From
Table~\ref{tab-line-strengths}, we find R$_{\textup{o}} = 7.7/41.1 = 0.19$ and
R$_{\textup{p}} = 14.4/10.0 = 1.4$, which again, as we noted earlier, puts
NGC\,1275 in the R$_{\textup{o}}$\,--\,R$_{\textup{p}}$ plane close to the
location of
NGC\,6240 (Fig.~\ref{fig-mouri-12}).  Starburst galaxies like NGC 253
and galactic photodissociation regions generally show R$_{\textup{p}} \le
1.0$.
Following this method one can also define a ratio, R$_{\textup{o}}$ =
H{$_2$}\,1--0\,S(3)/H{$_2$}\,1--0\,S(1), of two ortho lines which is again
independent of the
initial ortho-para ratio in the molecular clouds.  When placed in this
diagram (Fig.~\ref{fig-mouri-12}), NGC\,1275 falls near other powerful
AGNs like NGC\,1068.

The interpretation of these diagrams tells us that the excitation is
mostly ``thermal'', with a small component appearing as ``nonthermal
emission''.  The gas must then be quite dense, with
n$_{\textup{T}} \mbox{\lower0.6ex\hbox{$ \; \buildrel > \over \sim \;$}}\,
10^{5}$\,cm$^{-3}$.  The placement of NGC\,1275 near other powerful AGN
and merger galaxies suggests that both X-ray and shock mechanisms may
be at work creating the powerful H{$_2$} emission.  Hence, based on the
spatial extent of the H{$_2$} emission and its temperature, we propose
that there are two distinct sources of H{$_2$} excitation: the AGN
nucleus, which excites the vast majority of the radiation, and which
produces the compact emission core seen in our maps, and the extended
region of moderate star formation which excites the highly extended,
weak H{$_2$} envelope.

\subsection{H{$_2$} 2--1S(3) line strength}
\label{sec-2-1S3-strength}
The column density of the H{$_2$}\,2--1\,S(3) line is lower by a factor of 3
compared to what could be expected from weighted extrapolation of
 the 2--1\,S(1) and 2--1\,S(2) transitions.  Even if the ortho/para
ratio within the 2--1 transitions is not close to 3, which it was in
the 1--0 transitions, N$_{\textup{col}}$(2--1\,S(3)) is still a factor of
$> 2.5$ lower
compared to N$_{\textup{col}}$(2--1\,S(1)) and using T$_{\textup{ex}} =
2600$ \,K for the
two ortho states.  Inspecting the original data and the spectrum in
Fig.~\ref{fig-k-3secap-spectrum} did not show any abnormalities.
Since the chance for blending the line with a strong absorption can be
excluded for this part of the spectrum and the atmosphere, we argue
that this weakness of the H{$_2$}\,2--1\,S(3) line is real.  Inspecting
Fig.~\ref{fig-specfit-k-3} reveals that the 2--1\,S(1) line is also
weak outside the nucleus at a radius of $\approx 1$\,kpc.  It would be
visible if it was 2.5 time stronger or about half the strength
of the 1--0\,S(2) line, which is clearly present.

The only other galaxy, were a weak H{$_2$}\,2--1\,S(3) line has been reported
so far, is NGC\,6240 (Lester et al. \cite{Lester:88}), although only as an
upper
limit.  From their Table 3 one can estimate that the column density of
2--1\,S(3) line in NGC\,6240 is at least 3.5 times weaker than expected
from extrapolating other data, which is in good agreement with our
finding and suggest that there may be a common mechanism at work.  The
column densities in NGC\,6240 are only a factor of 2 higher compared
with NGC\,1275 an their derived temperature range is almost identical
with Fig.~\ref{fig-h2-exite-temp}.  It should be noted, however, that
the excitation conditions in both galaxies are not identical.  Draine
\& Woods (\cite{Draine:90}) have shown that in addition to X-ray heating,
UV pumping
plays a major role in NGC\,6240.  This is already indicated in
Fig.~\ref{fig-mouri-12} as well as and in Fig. 4 of Lester et
al. (\cite{Lester:88}):
The large offset between the line through the 1--0 transitions and the
2--1 data is different from NGC\,1275 (Fig.~\ref{fig-h2-exite-temp}).
Recently, Sugai et al. (\cite{Sugai:97}) reported that they were unable to
confirm the
weakness of the H{$_2$}\,2--1\,S(3) line in NGC\,6240 with R = 580 grism
spectroscopy.  Better quality data at higher spectral resolution is
certainly required to clarify the situation.

The most likely mechanism for the weakness of the H{$_2$}\,2--1\,S(3) line is
resonant fluorescent excitation of the upper level ($\nu$ = 2, J= 5)
by Ly$\alpha$ photons (Black \& van Dishoeck \cite{Black:87}).  The
Ly$\alpha$ photons needed for
this pumping process can easily be provided in an environment where a
notable fraction of the H{$_2$} is excited by X--ray heating
(Black \& van Dishoeck \cite{Black:87}).  Although detailed models of
resonant fluorescent
excitation of the $\nu = 2, J= 5$ level does not exist, the weakness
of the H{$_2$}\,2--1 \,S(3) line by itself already demonstrates that at least
in the nuclear region the excitation of the molecular hydrogen in
NGC\,1275 is dominated by the central AGN. This is in agreement with
our previous finding that the excitation environment in the nuclear
region of NGC\,1275 is much more dominated by X-ray heating than in
NGC\,6240 (Draine \& Woods \cite{Draine:90}).  NGC\,1275 may be the first
galaxy for
which a more complete model on resonant fluorescent excitation can be
computed.  Finding other galaxies with such a weak H{$_2$}\,2--1 \,S(3)
line (e.g. in NGC\,1068) would be important for constraining the process,
which may provide an additional tool for understanding the radiative
interaction between AGN and circumnuclear molecular material.

\begin{figure}
\resizebox{6cm}{8.5cm}{-}%\includegraphics{h1635_9a.eps}}
\resizebox{6cm}{8.5cm}{-}%\includegraphics{h1635_9b.eps}}

\caption{Excitation diagrams showing the relative contributions of
thermal and nonthermal excitation sources to H{$_2$} emission (styled
after Mouri 1994).
The position of a source in these diagrams is \emph{independent} of
the intrinsic ortho/para ratio in the molecular clouds.  In both
diagrams the regions occupied by shocks (Brand et al. 1989),
nonthermal (Black \& van Dishoeck 1987), thermal UV (Sternberg \&
Dalgarno 1989), and X-ray (Lepp \& McCray {\protect\cite{Lepp:83}},
Draine \& Woods {\protect\cite{Draine:90}})
excitation are outlined.  Data of NGC\,6240
and NGC\,1068 are taken from  Lester et al. (1988) and Oliva \&
Moorwood ({\protect\cite{Oliva:90}}).
The location of NGC\,1275 in these diagrams speaks for a
largely thermal excitation of \mbox{\lower0.6ex\hbox{$ \; \buildrel >
\over \sim \;$}}\,2000 \,K.
}

\label{fig-mouri-12}
\end{figure}

\section{ Br$\gamma$ emission}
\label{sec-brg}

\begin{figure}
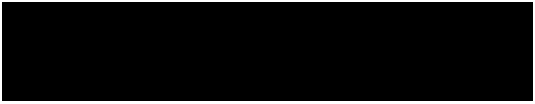

\resizebox{\hsize}{9cm}{-}%\includegraphics{h1635_10.eps}}
\caption{Line map of Br$\gamma$ showing clear asymmetrical extension.
	Solid line marks the radio jet position angle of 160$^\circ$.
	Contours are at 10, 20, 30\ldots\% of peak flux of
	$5.3 \times 10^{-15}\mbox{\, erg s$^{-1}$ cm$^{-2}$ arcsec$^{-2}$}$.}
\label{fig-brg-line-map}
\end{figure}

The Br$\gamma$ emission from NGC\,1275 is unusual in that it is highly
asymmetric both in its line profile
(Fig.~\ref{fig-k-3secap-spectrum}), and in its spatial distribution
(Fig.~\ref{fig-brg-line-map}).  The Br$\gamma$ emission is extended along
roughly the same line (PA =160$^\circ$) as the powerful radio jets
(Pedlar et al. \cite{Pedlar:90}), which suggests that the mechanism of its
ionization is more strongly influenced by the central engine than by
the surrounding stellar population.  This result correlates well with
H$\alpha$ and [NII] maps with visible band imaging spectroscopy
(Ferruit \& Pecontal \cite{Ferruit-tiger:94}).  In particular, the position
of the
extension in our maps correlates with an H$\alpha$ and [NII] emission
peak located $\approx 1\farcs2$ N, 0\farcs7 W of
the nucleus.  From the published H$\alpha$ maps, we estimate the H$\alpha$
flux in the hot spot to be $1.3\times 10^{-17}$ \, W m$^{-2}$
arcsec$^{-2}$.  Based on the
observed [SII] line ratios and assuming a temperature of
T$_{\textup{e}} =
10^{4}$\,K, (Ferruit \& Pecontal \cite{Ferruit-tiger:94}) derive a mean
electron
density of n$_{\textup{e}} = 40$ for this region, and a typical linewidth of
340\,km s$^{-1}$.

The intrinsic Br$\gamma$ linewidth at the nucleus of $980 \pm 30$ \,km
s$^{-1}$ is
unusually large for a pure star forming source.  However, the width is
not unexpected for emission from NLRs associated with AGNs and we
therfore safely assume here that the NLR contributes a substatial
fraction to the Br$\gamma$ emission.  The larger linewidth of Br$\gamma$
compared
to H$\alpha$ is probably a differential extinction effect.  The emission
was too faint to derive accurate Br$\gamma$ line profiles in regions 1\arcsec\,
to the NW and SE of the nucleus in order to test whether the excess
width and shape of the Br$\gamma$ line is due to expansion.  The total
Br$\gamma$
emission within the central 1\,kpc is $9.7\times 10^{-15}$ \,erg\,s$^{-1}$\,
cm$^{-2}$, which is a total luminosity of $1.5 \times 10^{6}$ \,L$_{\sun}$.
This corresponds to an ionization rate of $3.2 \times 10^{53}$ \,s$^{-1}$
or $2.0 \times 10^{9}$ \,L$_{\sun}$ for T$_{\textup{e}} = 7500$\,K, and can
be compared
with the value of $1.1\times 10^{54}$ \,s$^{-1}$ inferred from M82 over a
similar region from H53$\alpha$ recombination lines
(Puxley \cite{Puxley:91}).  From the ionization rate one can infer the total
Lyman continuum luminosity L$_{\textup{Lyc}}$ if one assumes OB type stars
as its
source.  L$_{\textup{Lyc}} = 2.0\times 10^{9}$\,L$_{\sun}$ yields about
1.5\% of the
L$_{\textup{FIR}}$ of NGC\,1275 ($1.5 \times 10^{11}$\,L$_{\sun}$) or
L$_{\textup{FIR}}$/
L$_{\textup{Lyc}} \le 75$.  L$_{\textup{Lyc}} = 2\times 10^{9}$\,L$_{\sun}$
is a lower limit to the
Lyc-luminosity in the IRAS beam.  The total bolometric luminosity of
such an evolving stellar cluster is typically about 10 - 20 times
higher than the Lyman continuum luminosity for typical assumptions on
the properties of such a region if it were an active star forming
region powered by massive stars (age $\sim$ a few 10$^{7}$\,y,
M$_{\textup{lower}} =
0.1$\,M$_{\sun}$, M$_{\textup{upper}} = 120$\,M$_{\sun}$, see Fig. 10 in
Krabbe et al. \cite{Krabbe:94}).
With that ratio L$_{\textup{Bol}}^{\textup{ionizing}} \sim 3.8 \times
10^{10}$L$_{\sun}$,
which is about 25\% of L$_{\textup{FIR}}$. Taken as face values these
numbers are
not very consistent with star formation but much more with what one
gets empirically, using above formulas for AGNs (Genzel et al.
\cite{Genzel:98}).

\section{The iron emission}
\subsection{Strength and morphology of the \mbox{[Fe\,{\sc ii}]
emission}}
\begin{figure}
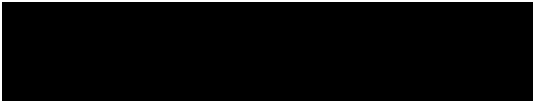

\resizebox{\hsize}{9cm}{-}%\includegraphics{h1635_11.eps}}
\caption{Line map of the FeII 1.644 \,$\mu$m\, line showing faint
extensions beyond
	the immediate nuclear area.
	Contours are at 1.25, 5, 10, 20, 30\ldots\% of peak flux of
	$3.0\times 10^{-14}$ \, erg s$^{-1}$ cm$^{-2}$ arcsec$^{-2}$.}
\label{fig-feii-line-map}
\end{figure}

The \mbox{[Fe\,{\sc ii}]} 1.644 \,$\mu$m\, emission from NGC\,1275 is very
strong
(Fig.~\ref{fig-h-3secap-spectrum}) and
centrally concentrated (Fig.~\ref{fig-feii-line-map}).  Other
prominent H and K band iron lines include the \mbox{[Fe\,{\sc ii}]} 1.534
\,$\mu$m,
\mbox{[Fe\,{\sc iii}]}\,
2.2178 \,$\mu$m, and \mbox{[Fe\,{\sc iii}]}\, 2.2420
\,$\mu$m, in a near blend with H{$_2$}\,2--1\,S(1).
The flux of the \mbox{[Fe\,{\sc ii}]} 1.644 \,$\mu$m\,
(Table~\ref{tab-line-strengths}) is
nearly as large as the \mbox{[Fe\,{\sc ii}]} 1.257 \,$\mu$m\, in an
8\arcsec\, aperture by
(Rudy et al. \cite{Rudy:93}).  The \mbox{[Fe\,{\sc ii}]} emission is much
stronger compared to other
starburst/Seyfert indicators than it is in most active galaxies.  For
example, in the central 1\,kpc of NGC\,1275, we find
\mbox{[Fe\,{\sc ii}]}($\lambda$1.644)/Br$\gamma$ = 6.3, whereas in roughly
similar sized
regions in the merger driven starburst NGC\,3256 \mbox{[Fe\,{\sc
ii}]}/Br$\gamma$ = 1.55
and in the starburst/Seyfert systems NGC\,4945 \mbox{[Fe\,{\sc
ii}]}/Br$\gamma$ = 1.4
(Moorwood \& Oliva \cite{Moorwood:94}) and NGC\,1068 \mbox{[Fe\,{\sc
ii}]/Br$\gamma$ =
2.5} (Thatte et al., in preparation; Blietz et al. \cite{Blietz:94}).
While Br$\gamma$ originates predominantly from photoionization of
\mbox{H\,{\sc ii}}\,
regions by OB stars, \mbox{[Fe\,{\sc ii}]} has an additional source.
Most of the gas phase Fe in the local ISM condenses onto grains, so
that the free Fe is often depleted by factors of 100 -- 1000.  If we
presume that the ionization characteristics of the central source in
NGC\,1275 are similar to those of NGC\,3256, NGC\,4945, and
NGC\,1068, then the unusual \mbox{[Fe\,{\sc ii}]}/Br$\gamma$ and
\mbox{[Fe\,{\sc ii}]}/\mbox{Pa$\beta$}\, ratios
(e.g. Rudy et al. \cite{Rudy:93}) would suggest that the iron depletion is
\mbox{\lower0.6ex\hbox{$ \; \buildrel < \over \sim
\;$}}\,10 in NGC\,1275, less severe than in most starburst and active
galaxies.  This could be due to an ambient radiation field which heats
the dust grains and tends to evaporate the Fe off of them.  Such a
scenario is supported by the relatively high derived dust temperature
within the central 300\,pc (e.g., Sect.\,\ref{sec-decomposition}).
Alternatively, the fact that the ionization state of the gas must
allow for plentiful Fe$^{+}$, which can then be further ionized by
photons with $E > 16.2$ eV, requires that the ionizing source of
radiation must be relatively cool, in order to produce low ionization
radiation which would produce the high \mbox{[Fe\,{\sc ii}]}/Br$\gamma$
line ratios.
However, fast J-shocks and X-ray heating can also produce copious
\mbox{[Fe\,{\sc ii}]} emission (Blietz et al. \cite{Blietz:94}).  In the
latter case, the X-rays
penetrate deep into molecular clouds where they create large
reservoirs of partially ionized gas.

\subsubsection{Profile and excitation of the \mbox{[Fe\,{\sc ii}]} emission}
\label{sec-fe-excitation}

\begin{figure}
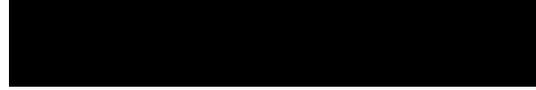

\resizebox{\hsize}{8cm}{-}%\includegraphics{h1635_12.eps}}
\caption{Profile of the FeII 1.644 \,$\mu$m\, line showing multicomponent
fit.  The
	solid dark line is the data, to which the best single component fit
	has a full width of 544 \,km s$^{-1}$. The two lower
	intensity profiles are the broad and narrow component required to
	fit the profile well.  The dotted line which closely follows the
	data is the sum of the broad ($1345\pm 120$\,km s$^{-1}$) and narrow
	($280\pm20$ \,km s$^{-1}$) components.}
\label{fig-feii-profile}
\end{figure}

The profile of the \mbox{[Fe\,{\sc ii}]} 1.644 \,$\mu$m\, line cannot be
well modeled as a
single Gaussian.  This is typical of Seyfert galaxies, which often
show narrow and broad components of the same emission feature.  We
therefore fit the observed profile using two Gaussians, a broad
component of width $1340 \pm 120$\,km s$^{-1}$ and relative amplitude of
$0.26$, and a narrow component of of width $280 \pm 20$\,km s$^{-1}$.  This
two component fit (Fig.~\ref{fig-feii-profile}), is excellent, and
suggests that $\approx 50\%$ of the \mbox{[Fe\,{\sc ii}]} emission arises
in the same
region as the Br$\gamma$, showing a similar linewidth, while the other half
of the \mbox{[Fe\,{\sc ii}]} arises in the same region as the H{$_2$},
where the
linewidths are again similar.  Hence we propose --- as in the case of
the H{$_2$} lines (Sect.\,\ref{sec-h2-excitation}) --- that the active nucleus
of NGC\,1275 is responsible for much of the excitation leading to
\mbox{[Fe\,{\sc ii}]}
emission.  Comparing this result with Blietz et al.  (1994), who have
found an excellent correlation between the \mbox{[Fe\,{\sc ii}]} emission
and the NLR
in NGC\,1068 suggests, that we might have a similar case here but
unresolved and at much greater distance.

The detailed line ratios expected from excitation via a central AGN
have been modeled by Hollenbach \& Maloney (\cite{Hollenbach:96}).  In
their model, the X-ray
energy absorbed per unit time per hydrogen nucleus in a cloud at
distance 100r$_{100}$ \,pc is H$_{\textup{X}} \sim 7\times 10^{-22}$
L$_{44}$r$_{100}^{-2}$N$_{22}^{-1}$ \, erg\,s$^{-1}$, where
$10^{44}$L$_{44}$ \,erg \,s$^{-1}$ is the AGN luminosity and
$10^{22}$N$_{22}$ \,cm$^{-2}$ is the
total column density.  The heating rate is $\approx 0.3$\,--\,0.4
H$_{\textup{X}}$, and the cooling is due to line emission from many
species, but
of particular diagnostic interest are the H{$_2$}\,1--0\,S(1) and
\mbox{[Fe\,{\sc ii}]}
1.64 \,$\mu$m\, lines.  Using our value of
N$_{\textup{H}_{2}^{\textup{hot}}} = 3\times
10^{16}$ \,cm$^{-2}$, a molecular abundance of H/H{$_2$} $\sim 10^{5}$, and
L$_{\textup{X}} \approx 2\times 10^{43}$ \,erg \,s$^{-1}$ (Prieto
\cite{Prieto:96}), and
integrating the Hollenbach \& Maloney results over our resolution beam
of 300\,pc, we can account for the observed \mbox{[Fe\,{\sc ii}]} to
H{$_2$}\,1--0\,S(1)
line ratio.  For all radii larger than our beam, the predicted ratio
is $\approx 1:1$.  This is close to our measured value of 1.5:1 in a
3\arcsec aperture (Table~\ref{tab-line-strengths}) since the
predicted ratio depends on the presumed depletion of Fe, which we
already know to be $\mbox{\lower0.6ex\hbox{$ \; \buildrel < \over \sim
\;$}}\, 10$, and which the model presumes to be
$\approx 30$ (Hollenbach, private communication).  We thus conclude that
the \mbox{[Fe\,{\sc ii}]} emission as well as \mbox{[Fe\,{\sc
ii}]}/H{$_2$}\,1--0\,S(1) can be explained by
an X--radiation field typically emitted by AGN.

\section{The CO$_{\textup{sp}}$ index}
\label{sec-co-index}

\begin{figure}
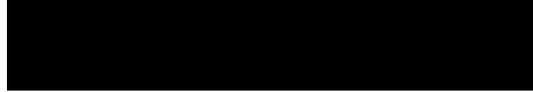

\resizebox{\hsize}{8.5cm}{-}%\includegraphics{h1635_13.eps}}
\caption{The azimuthally averaged CO$_{\textup{sp}}$\, index in NGC\,1275.
	The apparent decrease in CO$_{\textup{sp}}$\, toward the center is an
           artifact of continuum dilution due to hot dust and the scattered
light
          of the AGN. }
\label{fig-co-index}
\end{figure}

\begin{figure}
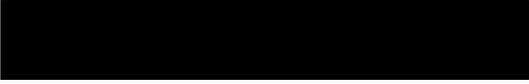

\resizebox{\hsize}{7.5cm}{-}%\includegraphics{h1635_14.eps}}
\caption{The black lines in the graph represent solutions
	of equation (\ref{eqn-co-dilution}). The inner area of the triangle
	illustrate the range of allowed values of the measured CO
	spectroscopic index CO$_{\textup{sp}}(\chi$) in NGC\,1275.
Constrains are the
	limits set by the error of the index itself, the range of likely
	values for the spectroscopic index  CO$_{\textup{sp}}$(0) in a
nondiluted stellar
	cluster, and the error in $\chi$ (see Fig. \,\ref{fig-co-index} and
	Sect. \,\ref{sec-decomposition}.)}
\label{fig-co-dilution-check}
\end{figure}

The CO$_{\textup{sp}}$ index of the underlying stellar population can be
diluted by
continuum emission not originating in stars.  If the non-stellar
continuum is $\chi$ times stronger than the stellar continuum, the
CO$_{\textup{sp}}$ index will be diminished according to
\begin{equation}
	~~~~~~~~~CO_{sp}(\chi) = -2.5\,log\,\frac{S_{0}+\chi}{1 + \chi}
	\label{eqn-co-dilution}
\end{equation}
where S$_{0} = 10^{-0.4\textup{CO}_{\textup{sp}}(0)}$ is the average of the
undiluted
rectified spectrum between 2.31 and 2.40 \,$\mu$m\, and
CO$_{\textup{sp}}$ is
the spectroscopic index. Details are given in ~Doyon et al.
(\cite{doyon-co:94}).
According to our model fitting, the non-stellar continuum in the
central 1\arcsec\, contributes $\mbox{\lower0.6ex\hbox{$ \; \buildrel >
\over \sim
\;$}}\, 90\%$ of the emission, reducing the
observed CO$_{\textup{sp}}$ from its intrinsic value of $\approx 0.3$ to
$\approx
0.03$, and and giving rise to the CO$_{\textup{sp}}$ hole at the nucleus of
NGC\,1275 (Fig.~\ref{fig-co-index}).  The range for the allowed
values of the intrinsic CO$_{\textup{sp}}$ is indicated in
Fig.~\ref{fig-co-dilution-check}.  The nuclear stellar cluster
may have a higher CO index than initially assumed, implying that
the K supergiants with CO$_{\textup{sp}}\approx 0.3$ instead of giants may be
the dominating population (Doyon et al. \cite{doyon-co:94}).  Including
this finding into a reiteration
on the continuum decomposition will not have any effect on the
results, since the temperature of these stellar types are not very
different.

Outside of the hole, the CO$_{\textup{sp}}$ of 0.17 is
typical of E/S0 galaxies (Frogel et al. \cite{Frogel:78}), confirming that
the stellar
population of NGC\,1275 is quite normal.  Furthermore CO$_{\textup{sp}} = 0.17$
corresponds to a \mbox{K5\,\sc III}\, star (Doyon et al.\cite{doyon-co:94},
appendix A), which
corroborates our spectral synthesis results for regions at R
\mbox{\lower0.6ex\hbox{$ \; \buildrel > \over \sim
\;$}}\, 3\,\arcsec.

Given that the light is dominated by K giants or supergiants, we try
to estimate the total nuclear stellar mass.  Using our result that the
emission enclosed in the central 3\arcsec\, (1\,kpc) of NGC\,1275 is
$\mbox{\lower0.6ex\hbox{$ \; \buildrel > \over \sim \;$}}\, 90\%$ due
to non-stellar sources, we correct the observed K magnitude of the
\emph{stellar component only} of m$_{\textup{K}}=11.8$ [in agreement
with other recent values m$_{\textup{K}}=11.9$ (Forbes et al.
\cite{Forbes:92})] to
m$_{\textup{K}}=14.3$.  Converting to absolute visual magnitudes
yields M$_{\textup{V}}^{\textup{K5\,III}}$ (NGC\,1275)\,=\,(V-K) +
m$_{\textup{K}} - 25 - $A$_{\textup{K}}-5 $ log D$_{\textup{Mpc}} =
-16.8$ using V-K= 3.24, A$_{\textup{K}} \sim 0.2$, and D = 68.8\,Mpc.
The mass of a K giant is 1.2 \,M$_{\sun}$, and its
M$_{\textup{V}}^{\textup{K5\,III}} = -0.2$ (Lang \cite{Lang:92}), which
implies $4.6 \times 10^{6}$ \,M$_{\sun}$ of \mbox{K5\,\sc III}\, stars
in the central 3\arcsec.

Assuming a Salpeter IMF with P(M)$\,\propto$\,M$^{-1.35}$ dM, the
mass fraction of the stars in the range 1.0 to 1.5 \,M$_{\sun}$ (roughly the
range of stellar masses which will become \mbox{K5\,\sc III}\, stars), for a
population bounded by 0.1 \,M$_{\sun}$ and 100 \,M$_{\sun}$ stars is
$(1.0^{-0.35} - 1.5^{-0.35})/(0.1^{-0.35} - 100^{-0.35}) = 0.065$, so
that \mbox{K5\,\sc III}\, stars make up about 7\% of the mass of a star
cluster.
Hence the total mass of stars in the central kpc is thus $7.1\times
10^{7}$ \,M$_{\sun}$. If the population is dominated by K supergiants, the
total stellar mass comes out to be the same value (mass range =
12 -- 14 \,M$_{\sun}$ M$_{\textup{V}}^{\textup{K5\,I}} = -5.0$).

We estimate the core radius r$_{0}$ of the nuclear stellar cluster to
be smaller than 500 \,pc.  Using this value as an upper limit and
knowing the stellar mass of the cluster, we can use $\rho_{0}=9
\sigma^{2}_{0} / 4 \pi G r^{2}_{0}$ (Eckart et al. \cite{Eckart:93}) to
estimate the
expected velocity dispersion: $\sigma = \sqrt{M_{\textup{cluster}}G / 9 r_{0}}
\mbox{\lower0.6ex\hbox{$ \; \buildrel > \over \sim
\;$}}\, 8.4$ km/s.  Even if the CO absorption had been detected on the
nucleus, such a small dispersion
had been beyond the capabilities of our instrument.

We conclude that the absence of CO bandheads
in the nuclear emission indicates that those bandheads are heavily
diluted by the emission of hot dust and power-law emission.

\section{Discussion}
Our spectral synthesis (Sect.\,\ref{sec-cont-emission}) shows that inside
of the central $300\,pc$, the NIR continuum emission is dominated by
the AGN and hot dust components.  Together these are
\mbox{\lower0.6ex\hbox{$ \; \buildrel > \over \sim \;$}}\, 10 times
more powerful than the stellar emission from the same region.  This
result is confirmed by the observed factor of
\mbox{\lower0.6ex\hbox{$ \; \buildrel > \over \sim \;$}}\, 10 continuum
CO$_{\textup{sp}}$ dilution (Sect.\,\ref{sec-co-index}) at the nucleus.
The nature of
the nuclear power source likely is a massive black hole (MBH) with
a surrounding molecular gas cloud torus or disk. The lack of strong CO
absorption bands indicates that massive star formation probably does
not play a very important role in the nucleus.

The intensity and morphology of the line emission suggest that more
than one mechanism is at work exciting the line emission.  In addition
to the AGN, shocks are also likely to play a large role.  Since the J
band \mbox{[Fe\,{\sc ii}]} lines are unusually strong, even for an AGN, it
requires a
substantial boost of the H{$_2$} emission to keep the [Fe\,{\sc
ii}]/H{$_2$}\,1--0\,S(1)
ratio near unity.  This is indeed observed in NGC\,1275, which has
unusually strong H{$_2$} emission compared to other Seyferts.
Furthermore, the spatial extension of the H{$_2$} emission
(Sect.\,\ref{sec-h2-distribution}) suggests that it may be driven by shocks
from a recent or ongoing merger event.  Such a picture is supported by
not only the observed isophotal shell structure (Hernquist \& Weil
\cite{Hernquist:92}),
but also by the extinction analysis of Norgaard-Nielsen et al.
(\cite{Norgaard-Nielsen:93}),
which shows that the HV system is at least partially \emph{moving
through} the main galaxy rather than completely in front of it.  Hence
the interaction of the HV and LV systems could give rise to the
required shocks which boost the H{$_2$} emission.  This process is
postulated to occur in the merging galaxy NGC\,6240
(van der Werf et al. \cite{vanderWerf:93}).  Among AGNs, strong
\mbox{[Fe\,{\sc ii}]} emission is not
unique to NGC\,1275.  In fact there are a large number of AGNs with
elevated \mbox{[Fe\,{\sc ii}]} emission (among them many IRAS luminous and
starburst
galaxies such as MRK\,231, MRK\,507, and PHL\,1092), which can be
attributed to violent star formation in a metal rich environment
(Fillipenko \& Terlevich \cite{Fillipenko:93}, Lipari et al.
\cite{Lipari:93}).  In the merger induced H{$_2$} boosting
scenario, the unusually high level of \mbox{[Fe\,{\sc ii}]} emission
observed from
NGC\,1275 would likely be due to a different effect, namely the
dissociation of Fe from interstellar grains via shock heating.  This
could produce an iron rich gas phase (Sect.\,\ref{sec-fe-excitation}),
which would then give rise to the very strong observed \mbox{[Fe\,{\sc
ii}]} emission.

The fact that the Br$\gamma$ emission is generally extended suggests the
presence of young stars and thus extended recent star formation.  If
the slight extension of the continuum is significant, it suggests
increased star formation along the jets, perhaps due to shocks, caused
by interactions with the surrounding medium.  Part of the Br$\gamma$
emission along the radio jet may be influenced by the radiation from
the nucleus and thus not created by star formation.  The extended
component of hot molecular hydrogen is probably shock excited as a
result of the star formation.

The existence (or non-existence) of ``cooling flows'' --- an inflow of
cool gas into a galaxy from its halo of hot x-ray emitting gas --- is
currently a topic of intense debate.  The evidence \emph{for} them is
that some galaxies which are surrounded by a halo of hot, X-ray
emitting gas show a dramatic central drop in the gas temperature.  As
the primary cooling mechanism of this gas is bremsstrahlung (emission
$\propto$\, n$^{2}$), and as the central temperature drop is accompanied
by a brightness increase, the gas must be both cooler and denser
(see, e.g. Sarazin et al. \cite{Sarazin:92} for a physical introduction and
review) .
If the gas is cooler and denser, then it must fall toward the center
of the galaxy.  In the case of NGC\,1275, the predicted infall amounts
to $\approx 200$ \,M$_{\sun}$\,yr$^{-1}$ (Fabian et al.
\cite{Fabian:81}), or $2 \times
10^{12}$ \,M$_{\sun}$ in $10^{10}$ years.  The evidence \emph{against}
cooling flows is that if such enormous quantities of gas really are
falling in, they must be detectable in some way, yet careful searches
have not revealed the gas in any of its expected states.  We know that
the gas cannot form stars with a normal IMF, because 200\,M$_{\sun}$ of
star formation would be 20 times more powerful than the prototypical
starburst galaxy, M\,82, and easily detected via color variations
(Cardiel et al. \cite{Cardiel:95}).

Our H{$_2$} data provide several evidences against the cooling flow
hypothosis.

\emph{i}) The spatial distribution of the hot molecular hydrogen
strongly peaks at the nucleus of NGC\,1275 on a scale $<$300pc.  If the
H{$_2$} emission was due to the cooling flow, one would have expexted a
much more extended distribution since the gas has to fall through all
of the galaxy.  The weakness of the H{$_2$}\,2--1\,S(3) line (see
Sect.\,\ref{sec-2-1S3-strength}) emphasises the close relation of the hot
H{$_2$}
with the nucleus.

\emph{ii}) The temperature variation within the hot H{$_2$} is small.
Between 0\,K and about 20000\,K upper level energy the temperature only
varies by about 1500\,K. A cooling hot gas cloud is likely to show a
much larger span of temperature compared to what our data show.

\emph{iii}) The temperature of the hot H{$_2$} is rather low: between
1500\,K and 3000 K. There is no evidence of a very hot
(\mbox{\lower0.6ex\hbox{$ \; \buildrel > \over \sim \;$}}\, 3000 K) H{$_2$}
component
which can be expected for a cooling hot gas mass.

\emph{iv} The line emission of the hot H{$_2$} can be explained consistently
with mostly thermal excitation (e.g. in a postshock environment) with
a smaller component of X--ray heating. A ``cooling flow'' scenario
does not need to be invoked to explain the line emission of molecular
hydrogen in NGC\,1275.

Several groups have argued that the gas cannot have collected in the
form of molecular hydrogen (H{$_2$}), because searches for CO emission
associated with H{$_2$} have revealed only very limited quantities of H{$_2$}
gas, although NGC\,1275 is unusual in that it shows more CO than most
cooling flow galaxies (Lazareff et al. \cite{Lazareff:89}, Inoue et al
\cite{Inoue:96}, O'Dea et al. \cite{Odea:94}).
However, since the temperature regime is very different from the cold
molecular clouds, the CO $\rightarrow$ H{$_2$} conversion factor is
certainly different and may even not be applicable in this context.

The gas could have been collected as
atomic hydrogen. However, only very limited quantities have
been detected, certainly much less than would be expected from an
infall of several hundred solar masses per year over cosmic timescales
(Haschick et al. \cite{Haschick:82}).

One of the last places in which the purported
inflow could be hiding is in stars with a very unusual kind of IMF,
one that is biased toward producing low mass stars, so that the star
formation rate as determined by the total number of ionizing photons
(these are produced only by massive stars) does not reflect the total
amount of mass involved in the star formation process.  A population
of such low mass stars should have a somewhat lower CO$_{\textup{sp}}$\,
index than a
normal population, because it will have fewer luminous giants and
supergiants which are the largest contributors to the CO absorption
lines.  Our NIR spectroscopy allows us to analyze the stellar
population of regions 1\,kpc from the nucleus via the CO$_{\textup{sp}}$\,
index,
where we find a CO$_{\textup{sp}} = 0.17$.  This can be compared with
theoretical
predictions for a population of very low mass stars (see Fig.
20 in Kroupa \& Gilmore \cite{Kroupa:94}), whose expected value of
CO$_{\textup{sp}}$ is between 0.07 and
0.16 for cooling flow rates of 158 \,M$_{\sun}$ \,yr$^{-1}$.  For higher mass
deposition rates, the expected CO$_{\textup{sp}}$ decreases, so that for a
rate of
316\,M$_{\sun}$\, yr$^{-1}$ the expected CO$_{\textup{sp}}$ lies between
0.0 and 0.1.
Hence we cannot completely eliminate the existence of a low mass star
population for a cooling flow of 200\,M$_{\sun}$ \,yr$^{-1}$ as is postulated
for NGC\,1275, but the observed CO$_{\textup{sp}}$ index is at the extreme
limit of
theoretically acceptable values.

\section{Summary}

We observed the central 1\,kpc of NGC\,1275 with the MPE 3D
imaging spectrograph.  We find that the central $\approx 300$\,pc\, are
strongly dominated by emission from an AGN
and from hot dust: There is no evidence of a nuclear stellar continuum.
In the regions at a distance of $\approx 1$\,kpc from the nucleus, the
situation is completely reversed, and the emission is totally
dominated by a stellar population whose spectra is consistent with a
normal, old ($>10^{8} - 10^9$ y)  population.  In this region, we find
no evidence to support the
thesis that substantial quantities of infalling ``cooling flow'' gas
has been transformed into stars with an IMF biased toward low masses,
although we cannot strictly rule out a low mass population because the
purported flow may be disturbed in the central 10\,kpc of the galaxy.
Within the central $\sim 500\,$pc, the emission line ratios of many
important diagnostic lines from the NIR regime are consistent with
excitation from a combination of AGN emission and shock excitation.
The shocks may be due to previous mergers, or to the early phases of a
merger between the HV and LV systems.

\begin{acknowledgements}
We thank the 3D team for their enthusiastic support during the
observations.

\end{acknowledgements}

\appendix
\section{Spectral Fitting of the continuum emission}
The total intrinsic continuum emission at each wavelength
$\lambda$ is
\begin{equation}
	I_{\lambda}^{\textup{INT}}=
	\sum_{\textup{i}}c_{\textup{i}}I_{\lambda}^{\textup{i}}
	\label{eqn-sum-cont-component} ~~~~with~~~~
	\sum_{\textup{i}}c_{\textup{i}} = 1 \mbox{.}
	\label{eqn-sum-unity}
\end{equation}
In active and starburst galaxies, we identify four sources of continuum
emission: stellar
($I_{\lambda}^{\textup{STAR}}$), AGN
power-law($I_{\lambda}^{\textup{P}}$) and hot dust
($I_{\lambda}^{\textup{HD}}$), and free-free
($I_{\lambda}^{\textup{FF}}$).
The intrinsic emission of
\begin{equation}
	I_{\lambda}^{\textup{INT}}=
	c_{\textup{S}}\mbox{$I_{\lambda}^{\textup{STAR}}$}\, +
	c_{\textup{PL}}I_{\lambda}^{\textup{P}}\,  +
c_{\textup{HD}}I_{\lambda}^{\textup{HD}}\, +
	c_{\textup{FF}}I_{\lambda}^{\textup{FF}}
	\label{eqn-total-emission}
\end{equation}
is then extincted according to
\begin{equation}
	I_{\lambda}^{\textup{SYN}}=
	\cases{
	     I_{\lambda}^{\textup{INT}} e^{-g\tau_{\lambda}}
e^{-\tau_{\lambda}}, & \mbox{Screen}\cr
	     I_{\lambda}^{INT}  e^{-g\tau_{\lambda}} (1 -
e^{-\tau_{\lambda}})/\tau_{\lambda}, & \mbox{Mix}.\cr
                 }
	\label{eqn-extinct-cases}
\end{equation}
to obtain the synthetic spectrum.
The extinction has two parts, one due to dust within our Galaxy
($g\tau_{\lambda}$), and one due to dust within the source
galaxy ($\tau_{\lambda}$).  In our model, the latter can have either a simple
screen geometry or a mixed stars/gas geometry.  In all cases, the
extinction is taken to vary as $\lambda^{-1.85}$ (Landini et al.
\cite{Landini:84}),
and is fixed to be A$_{\textup{K}} = 0.112$ A$_{\textup{V}}$ at $\lambda =
2.2\,\mu$m\, (Rieke \& Lebofsky \cite{Rieke:85}).

When comparing $I_{\lambda}^{\textup{SYN}}$ with the observed spectrum,
$I_{\lambda}^{\textup{OBS}}$, we avoid emission lines due to non-continuum
sources (e.g. shocked ISM gas, photoionized gas clouds), and define
our goodness of fit parameter, $\chi^{2}$, such that
 \begin{equation}
 	\chi^{2} = \frac{1}{N-1}\sum_{\lambda \neq line}
 	\frac{(I_{\lambda}^{\textup{SYN}}- I_{\lambda}^{\textup{OBS}})^{2}}
{\sigma_{\lambda}^{2}}
 	\label{eqn-chisqr}
 \end{equation}
 where $N$ is the number of measured spectral points which contain no
 line emission or absorption from non-continuum sources and
 $\sigma_{\lambda}$ is the local noise.

To model the stellar component, we have resampled the R$\approx 3000$
normalized K band spectra of Kleinmann \& Hall (\cite{Kleinmann:86}) at the 3D
resolution of R = 670 onto our $0.001 \,\mu$m \,pix$^{-1}$ data format
(F\"orster et al. \cite{Forster:96}), and then multiplied them by a
Rayleigh-Jeans
power law of $\lambda^{-3.94}$ (Doyon et al. \cite{doyon-co:94})
corresponding to
an A0V star to reconstruct the true stellar spectrum.

The intrinsic spectrum of the hot dust of temperature T$_{\textup{d}}$, is
taken to be a black body multiplied by a power-law emissivity of the
form $\epsilon_{\textup{dust}} \propto \lambda^{-1}$.  Hence,
$I_{\lambda}^{\textup{HD}}\,=
B_{\lambda}(T_{\textup{d}})\lambda^{-1}$, where
B$_{\lambda}(T_{\textup{d}})$ is the
black body function.

The behavior of free-free emission is given by $I_{\nu} \propto
e^{\textup{-h}\nu\textup{/kT}}$ where T is the plasma temperature (normally
$> 5000$\,K).
We convert to wavelength units, and find $I_{\lambda} \propto
\lambda^{-2} e^{\textup{-hc/}\lambda \textup{kT}}$ which we adopt as our
functional form
for free-free emission.

\newpage
%\section{BIBLIOGRAPHY}
%\nocite{*}
%\bibliographystyle{aabib99}
%\bibliography{aamnem99,n1275}

\listofobjects
\end{document}